\documentclass[a4paper,12pt]{article}
\usepackage{epsfig,epsf}

\usepackage{citesort}
 \normalsize

\newcommand{\bea}{\begin{eqnarray}}
\newcommand{\eea}{\end{eqnarray}}

\begin{document}

\def\lsim{\raise0.3ex\hbox{$\;<$\kern-0.75em\raise-1.1ex\hbox{$\sim\;$}}} 
\def\gsim{\raise0.3ex\hbox{$\;>$\kern-0.75em\raise-1.1ex\hbox{$\sim\;$}}}

\def\dofigure#1#2{\centerline{\epsfxsize=#1\epsfig{file=#2, width=7cm, 
height=4cm, angle=0}}}
\def\dofigureb#1#2{\centerline{\epsfxsize=#1\epsfig{file=#2, width=7cm, 
height=10cm, angle=-90}}}
%

%
\begin{titlepage}
\vspace*{-2cm}
\begin{flushright}
CERN-PH-TH/2007-033\\
HIP-2007-04/TH
\end{flushright}

{\Large
\begin{center}
{\bf Testing PVLAS axions with resonant photon splitting}
\end{center}
}
\vspace{.5cm}

\begin{center}
Emidio Gabrielli $^{a}$ and Massimo Giovannini $^{b,c}$
\\[5mm]
\textit{ a) Helsinki Institute of Physics,
P.O.B. 64, 00014 University of  Helsinki, Finland }
\\
\textit{ b) Centro ``Enrico Fermi'',
Via Panisperna 89/A, 00184 Rome, Italy \\
c) Department of Physics, Theory Division, CERN, 1211 Geneva 23, Switzerland }
\vspace{1.5cm}
\begin{abstract}

The photon splitting  
$\gamma \to \gamma \gamma$ in a time-independent and inhomogeneous
magnetized background is considered when
neutral and ultra-light spin-0 particles are coupled to two-photons.
Depending on the inhomogeneity scale of the 
external field, resonant photon splitting can occur.
If an optical laser crosses a magnetic field
of few Tesla with typical inhomogeneity scale of the order of the meter, 
a potentially observable rate of photon splittings is expected for the PVLAS range 
of couplings and masses.
\end{abstract}

\vspace{1cm} 

{\it This paper is dedicated to Emilio Zavattini}
\end{center}
\end{titlepage}

Ultra-Light scalar/pseudoscalar 
particles have escaped, so far, direct detection. There are, however,
theoretical reasons to believe in their possible existence.
For instance, they are naturally predicted by diverse theories beyond the standard model 
endowed with spontaneously broken (global) continuous symmetries.
Because of the lack of any direct detection, these particles 
must be very weakly coupled to standard matter fields.
A pivotal example along this direction is  the (pseudo-scalar) Nambu-Goldstone 
boson associated with the spontaneously broken Peccei-Quinn symmetry \cite{PQ}.
The mass of this particle, customarily called axion \cite{WW}, is believed to be in the meV range \cite{sikivie1,axion_astro}.

As far as the electromagnetic interactions are concerned, the Lagrangian densities for an ultra-light pseudo-scalar 
($\varphi_{P}$) or scalar ($\varphi_{S}$) field  can be parametrized, for the purposes of this discussion, by two different 
couplings, i.e. $\Lambda$ and $\tilde{\Lambda}$: 
\bea
L_{P}= -\frac{1}{4\Lambda}\, F_{\mu\nu} \tilde{F}^{\mu\nu}\, \varphi_P\, ,\qquad 
L_{S}= -\frac{1}{4\tilde{\Lambda}}\, F_{\mu\nu} F^{\mu\nu}\, \varphi_S\, ,
\label{LeffP}
\eea
where 
$F_{\mu\nu}$ and $\tilde{F}^{\mu\nu}$ are, respectively, the Maxwell field strength
and its dual.

Astrophysical constraints \cite{axion_astro} demand that
the axionic coupling $\Lambda$ should be of the order of 
$10^{10}-10^{11}$ GeV, implying, together with the smallness of the corresponding mass, that 
the axion is (almost) stable in comparison with the age of the Universe. 
The axion would then be a potential dark matter candidate.

\begin{figure}[tpb]
\dofigure{3.1in}{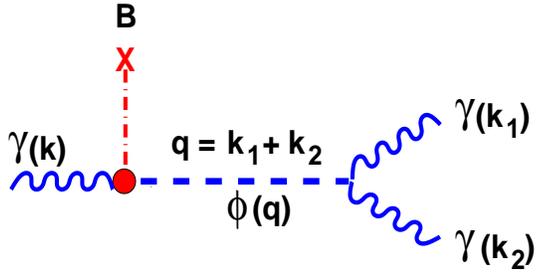}
\caption{\small The Feynman diagram describing the contribution  to the photon splitting
amplitude in an external magnetic field. The spin-0 particle $(\phi)$ 
is coupled to two-photons 
as in Eq. (\ref{LeffP}).  The dashed horizontal line indicates the propagator of the 
spin-0 particle, and the vertical dot-dashed line stands for the
external magnetic field insertion. The bubble represents the magnetic
form factor; $k$, $q$, and $k_{1,2}$ denote, respectively, the initial, intermediate 
and final four-momenta.}
\label{Figure}
\end{figure}

Recently, the PVLAS experiment \cite{PVLAS},
has reported the first evidence of a rotation of the polarization plane 
of light propagating through a magnetic field. According to the 
standard treatment \cite{MPZ},
these results would imply, if confirmed, the presence of a very light 
pseudoscalar particle (axion)  whose inferred mass $m$ and coupling $\Lambda$ will 
be, respectively,  ${\mathcal O}({\rm meV})$ and ${\mathcal O}(10^{6}\, {\rm GeV})$.
The purpose of the present paper is to test this claim by considering a 
complementary effect that has not received, so far, specific attention.
Our logic can be summarized, in short, by Fig. \ref{Figure}. 
Suppose that a laser beam 
passes through an inhomogeneous magnetic field. Now,  if the 
magnetic field can absorb momentum in a continuous manner, 
the (off-shell) axions can be reconverted into photons. In fact, 
the two-photon couplings in Eq. (\ref{LeffP}),
could also induce another indirect effect which is 
the photon splitting $\gamma\to \gamma\gamma$ in an external (time-independent) 
magnetic field. 
In a QED framework \cite{adler}, the rate of photon splitting is suppressed by 
$({\rm B/B^{\rm QED}_{\rm cr}})^6$, 
where ${\rm B}^{\rm QED}_{\rm cr}=m_e^2/e \simeq 4.4 \times 10^{9}$ Tesla.
Owing to the largeness of ${\rm B}^{\rm QED}_{\rm crit}$ the resulting effect is extremely 
minute for typical laboratory (i.e. ${\mathcal O}({\rm Tesla})$) magnetic fields.
The question we ought to address is therefore rather simple: how many 
photon splitting events are expected in the situation described by 
Fig. \ref{Figure}? 
If the magnetic field would be completely static and homogeneous 
the answer to this question would be only academic since,
due to the translational invariance of the full system, the process 
could only proceed by taking into account photon dispersion 
effects \cite{ghr}. Besides that, the process would also be suppressed by two 
extra powers of $\Lambda$.

If, on the contrary, the (classical) magnetic field is inhomogeneous, 
it is plausible to expect that the momentum of the photon beam could be partially absorbed in a continuous manner. 
In the latter case
 the answer to the aforementioned question is different since possible resonant effects must be taken into account.
Indeed there are two relevant physical scales in the problem at hand:
  $L$ (i.e.  the magnetic inhomogeneity scale of the external background field) and $\overline{p}$ (i.e. the momentum 
absorbed by the magnetic field at the resonance). If 
$L \sim {\mathcal O} (1/\overline{p})$ the photon splitting production 
rate is said to be resonantly amplified.   Amusingly enough the typical length 
scale $L$ turns out to be of the order of the meter for the PVLAS range of
 masses \cite{PVLAS} and taking, as incident beam, an optical laser.
While inhomogeneous magnetic fields have been considered in order to 
produce real (i.e. on-shell) 
axions by Primakoff effect \cite{sikivie1,Prim} (see also \cite{axion_exp1,BFRT}), 
their possible relevance for photon splitting has not been taken into account so far. 
One of the results of the present paper is that 
 the PVLAS axions should  also produce an observable excess of photon-splitting 
events in comparison with the (minute) QED background previously mentioned.
Consider therefore the process illustrated in Fig. \ref{Figure}, i.e. 
$\gamma (k)\to \gamma(k_1) \, \gamma(k_2)\,$,
where $k$ and $k_{1,2}$ are, respectively, the four momenta of the initial and of the final 
photons. Suppose also that the photons exchange momentum with the external magnetic field (represented 
by a cross in Fig.\ref{Figure}) which is assumed to be static (i.e. time-independent) 
but spatially inhomogeneous. Consequently,
while the total energy of the reaction is conserved (owing to the time-independence of the external field), the 
three-momentum can be absorbed by the (inhomogeneous) external field. 

Defining the four momenta of the final photons as
 $k_{1} =(\omega_{1},\vec{k}_{1})$ and $k_2=(\omega_{2},\vec{k}_{2})$,
the differential cross section for the scattering of a photon by
a time-independent external field is:
\bea
d\sigma = 
\, \frac{1}{2}\,
\sum_{\rm pol}|{\rm M}(\gamma\to \gamma\gamma)|^2
\,(2\pi)\delta(\omega-\omega_1-\omega_2)\, 
\frac{1}{2\omega}
\frac{d^3 \vec{k}_1}{2\omega_1 (2\pi)^3}
\frac{d^3 \vec{k}_2}{2\omega_2 (2\pi)^3},
\label{sigma}
\eea
where  ${\rm M}(\gamma\to \gamma\gamma)$ is the matrix element of the
process and the $1/2$-factor arises since the photons of the final state are indistinguishable. According to Fig. \ref{Figure} the scattering amplitude 
for  $\gamma\to \gamma\gamma$ in a magnetic field can be written as 
\bea
{\rm M}(\gamma\to \gamma\gamma)&=&
\frac{\sin{\theta}}
{\Lambda}\omega \, \hat{{\rm B}}(\vec{p})\,\frac{ i}{(k_1+k_2)^2-m^2+
i\,m\Gamma} V_{\gamma\gamma}(k_1,k_2)\, ,
\label{M}
\eea
where $m$ and $\Gamma$ are, respectively, the mass and total width of the exchanged boson;
moreover  $\theta$ is the angle
between the direction of the external magnetic field ${\rm \vec{B}}$ and 
the incoming photon three-momentum $\vec{k}$.
The term  $\hat{{\rm B}}(\vec{p})$ is the 
Fourier transform of the projection of $\vec{{\rm B}}(\vec{x})$
along the polarization vector of the incoming photon, namely
$
\hat{{\rm B}}(\vec{p}) 
= \int |\vec{n}\cdot \vec{{\rm B}}(\vec{x})|\, e^{i\,\vec{p}\cdot \vec{x}}d^3x
$, where $\vec{p}=\vec{k}-\vec{k}_1-\vec{k}_2\,$ is the momentum absorbed 
by the external field and $\vec{n}$ is the unit vector
parallel to the direction of the incoming  polarization. In Eq.(\ref{M}), 
the term
\bea
V_{\gamma\gamma}(k_1,k_2)=\frac{1}{\Lambda}
\left(k_1^{\mu}\, \varepsilon_1^{\nu}\, 
k_2^{\alpha}\, \varepsilon_2^{\beta}\right)\epsilon_{\mu\nu\alpha\beta}\, ,
\label{Vertex}
\eea
represents the boson-$\gamma\gamma$ vertex contribution  where 
$\varepsilon_1^{\mu}(k_1)$ and $\varepsilon_2^{\mu}(k_2)$ are 
the polarization vectors of the two final photons and 
$\epsilon_{\mu\nu\alpha\beta}$ is the total antisymmetric tensor. 

Inserting Eqs.(\ref{M}) and (\ref{Vertex}) into Eq.(\ref{sigma})
the differential cross-section becomes 
\bea
d \sigma&=&
\frac{\omega}{8\Lambda^4}
\frac{(k_1+k_2)^4\, \sin^2{\theta}}{((k_1+k_2)^2-m^2)^2+\Gamma^2m^2}
|\hat{\rm B}(\vec{p})|^2\, 
\nonumber\\
&\times&
 (2\pi)\delta(\omega-\omega_1-\omega_2)\, 
\frac{d^3 \vec{k}_1}{2\omega_1 (2\pi)^3}
\frac{d^3 \vec{k}_2}{2\omega_2 (2\pi)^3}.
\label{sigma1}
\eea
As previously mentioned, for the reaction to proceed, the 
three-momentum $\vec{p}$  must be absorbed by the external magnetic field. 
In fact, if the external field is fully homogeneous not only in time but also in space, we will
have that, in Fourier space,  
$|\hat{{\rm B}}(\vec{p})|\propto \delta^{(3)}(\vec{k}-\vec{k}_1-\vec{k}_2)$.
This occurrence demands that Eq.(\ref{sigma1}) is proportional to 
$(k_1+k_2)^2\, \delta^{(4)}(k-k_1-k_2)$. Then the total cross section vanishes after integrating over the total phase space.

Naively, if the magnetic field has a finite extension of 
order $L$ and vanishes outside of it,
the translational invariance is broken and the external 
field could easily absorb momentum of magnitude $\sim 1/L$.
To model this situation, consider, as an example, a magnetic field 
configuration with Cartesian components
$\vec{{\rm B}}(\vec{x}) =({\rm B}_x(\vec{x}),0,{\rm B}_z(\vec{x}))$, 
where 
\bea
{\rm B}_z(\vec{x})={\rm B}\, \exp\left(-\frac{x^2}{L^2}
-\frac{y^2 + z^2}{R^2}\right),\qquad
{\rm B}_x(\vec{x})=\frac{2z}{R^2}\,\int_{-\infty}^{x}\, d x \, 
{\rm B}_z(\vec{x})\,.
\label{Bshape}
\eea
It is easy to check that 
$\vec{{\rm B}}(\vec{x})$ is indeed solenoidal.
By choosing the three-momentum $\vec{k}$ 
of the incoming photon along the $\hat{x}$-axis, 
 only the $\hat{z}$ component of the magnetic field will enter Eq. (\ref{sigma1}) where
$|\hat{{\rm B}}(\vec{p})|\to |\int d^3 x {\rm B}_z(\vec{x})|$.

Let us then compute our main observable, i.e. the number of photon splitting events taking place 
 inside the magnetic field. This quantity, denoted in what follows by 
 $N_{\gamma\gamma}$, is simply 
 the product of  the cross section $\sigma$ (obtainable from Eq.(\ref{sigma1})) 
times the flux of incoming photons $\Phi=N_{\gamma}(t)/A$ , i.e. 
$N_{\gamma\gamma }(t)=\sigma \Phi$.
The quantity $N_{\gamma}(t)$ is the number of incoming  photons per unit 
time crossing the magnetic field and $A=\pi R^2$ 
is the surface spanned by the magnetic field on the 
the plane orthogonal to the direction of the
incoming photon momentum (in our frame the $\hat{y}-\hat{z}$ plane).
Inserting now Eq.(\ref{Bshape}) into Eq.(\ref{sigma1}),
the differential number of photon splitting events per unit time is given by 
$dN_{\gamma\gamma}(t)=
N_{\gamma}(t)\, d{\rm P}(\gamma\to\gamma\gamma)
$, where 
\bea
d{\rm P}(\gamma\to\gamma\gamma)
&=&\frac{\omega}{32\Lambda^4} {\rm B}^2
\frac{(k_1+k_2)^4}{((k_1+k_2)^2-m^2)^2+\Gamma^2m^2}
\,\pi\, L^2\,\exp{\left(-\frac{p^2 L^2}{2}\right)}\, \times
\nonumber\\
&& (2\pi)^3\delta(\omega-\omega_1-\omega_2)\, 
\Delta_{R}(\frac{k_{1y}+k_{2y}}{\sqrt{2}})\, \Delta_{R}(
\frac{k_{1z}+k_{2z}}{\sqrt{2}})\, \times
\nonumber\\
&&
\frac{d^3 \vec{k}_1}{2\omega_1 (2\pi)^3}
\frac{d^3 \vec{k}_2}{2\omega_2 (2\pi)^3}\, ,
\label{P2}
\eea
is the differential probability of $\gamma$ going to $\gamma\gamma$.
In Eq. (\ref{P2}),  the absorbed momentum is $p\equiv \omega-k_{1x}-k_{2x}$.
 The quantity
$\Delta_{L}(p)$ is nothing but
$
\Delta_L(p)\equiv (L/\sqrt{\pi}) \, 
\exp{\left( - L^2\, p^2\right)}$, which is a well known representation 
of the Dirac-delta function $\delta(p)$ in the limit
$\lim_{L\to \infty} \Delta_L(p)=\delta(p)$. To simplify the problem,
the shape of the magnetic field  can be chosen in such a way that 
$R\to \infty$ while  $L$ stays finite: in this  way the total momentum will be absorbed only along the initial beam direction $\hat{x}$. In practice this means that
$R \gg L$, i.e. $R$ much larger than the characteristic inhomogeneity scale along 
the $\hat{x}$ direction.  

After using the Dirac-delta functions for the partial integrations a change of variables allows to write the total probability as 
\bea
 {\rm P}(\gamma\to\gamma\gamma)
=\frac{{\rm B}^2}{256 \pi \Lambda^4}
[ {\mathcal I}_{1}(\omega) + {\mathcal I}_{2}(\omega)]\, ,
\label{Pgg}
\eea
where the following integrals have been introduced
\bea
{\mathcal I}_{1}(\omega) &=& \omega\,\int_{0}^{{\frac{\omega}{2}}}\, d\omega_1
\int_{0}^{4\omega_1(\omega-\omega_1)}\, d\mu^2
\frac{F(\mu^2)}{\left[(\mu^2-m^2)^2+\Gamma^2m^2\right]}\,,
\label{int1}\\
{\mathcal I}_{2}(\omega) &=& \omega
\,\int_{\frac{\omega}{2}}^{\omega}\, d\omega_1
\int_{0}^{2\omega(\omega-\omega_1)}\, d\mu^2
\frac{F(\mu^2)}{\left[(\mu^2-m^2)^2+\Gamma^2m^2\right]}\,, 
\label{int2}\\
F(\mu^2)&=&\,
\frac{\mu^4}{\omega^2-\mu^2}
\,\,  L^2\, 
e^{-L^2\,p^2/2}
,\qquad p=\omega(1-\sqrt{1-\mu^2/\omega^2}\, ).
\label{def}
\eea
The relevant physical regime for the present purposes is the one where $\omega > m$.
This limit is realized, for instance, when the incident photon beam is in the optical 
and the scalar mass in the PVLAS range.  If  $ m/\omega \ll 1$, 
the Breit-Wigner distribution can be easily 
integrated in the thin-resonance limit 
(since $ \Gamma = m^3/(64 \pi \Lambda^2)\ll m$), 
and, from Eq. (\ref{Pgg}) the result is simply 
\bea
{\rm P}(\gamma\to\gamma\gamma)&=&
\frac{{\rm B}^2\, \omega^2}{m^4\, \Lambda^2}  \, \frac{\pi^{3/2}}{2\sqrt{2}}
\,
{\mathcal F}(\chi) \,
{\rm BR}(\varphi_{P}\to \gamma\gamma)\, \left(1+\, {\cal O}(\frac{\Gamma}{m})
\, \right),
\label{K1}\\
{\mathcal F}(\chi) &=& \sqrt{\frac{2}{\pi}}\, \chi^2 e^{-\chi^2/2},
\qquad \chi = \overline{p}\, L,
\label{K1a}
\eea
where ${\rm BR}(\varphi_{P}\to \gamma\gamma)=m^3/(64\,\pi \Lambda^2\, \Gamma)$ 
is the branching ratio of the pseudo-scalar decay in $\gamma\gamma$
and terms of order ${\cal O}(m^2/\omega^2)$ have been neglected.
Furthermore, in Eq. (\ref{K1a}) $\overline{p}$
is the absorbed  momentum at the resonance, i.e.
\bea
\overline{p} = \omega\biggl( 1 - \sqrt{1 - 
\frac{m^2}{\omega^2}}\biggr)\, .
\label{RESP}
\eea
Note that ${\mathcal F}(\chi)$, being dimensionless and normalized to $1$, 
measures 
the suppression of $P(\gamma\to \gamma\gamma)$ when the magnetic inhomogeneity 
scale $L$ is larger than $1/\overline{p}$. 
The limit  $\omega \gg m$
is naturally implemented for the case of PVLAS axions when 
$\omega$ is the energy of the (optical) photon beam, and in this case
$\overline{p}\simeq m^2/(2 \omega)$.
Notice that the leading term of the integrals in Eq.(\ref{Pgg}) 
is proportional to $1/\Gamma$, enhancing the suppression induced
by the $1/\Lambda^4$ coupling. 
In particular, when the pseudo-scalar has only the decay channel in two-photon,
${\rm BR}(\varphi_{P}\to \gamma\gamma)=1$, Eq. (\ref{K1}) reproduces
the probability of photon conversion in on-shell axions \cite{sikivie1}.
Such an occurrence is a direct consequence of the Breit-Wigner distribution 
in Eq.(\ref{sigma1}).  It is relevant to mention that if $\omega < m$ (which is opposite to the 
limit where Eq. (\ref{K1}) has been derived), the region of the pole in the integrand of Eqs. (\ref{int1}) and (\ref{int2}) 
does not contribute to the total integral. Consequently, the total probability is simply suppressed as ${\rm B}^2/\Lambda^4$.

Generalizations of the result (\ref{K1}) to the scalar case are possible. In the scalar case 
the boson-$\gamma\gamma$ vertex of Eq. (\ref{Vertex}) is different and the coupling will be dictated by $\tilde{\Lambda}$. Consequently,
the relevant geometric set-up will be the one where the incoming photon polarization 
is orthogonal (and not parallel) to the orientation of the external magnetic field. 
Moreover, the dependence of the branching ratio for the
decay in $\gamma\gamma$ upon the mass, couplings and width of the scalar particle 
has the same analytical expression as in the pseudo-scalar case.
With these caveats the numerical results 
are the same if $\Lambda=\tilde{\Lambda}$.

According to Eq. (\ref{K1a}), the resonant effect in Eqs. (\ref{K1}) 
is achieved when $\chi =\sqrt{2}$ corresponding to the maximum of 
 ${\mathcal F}(\chi)$ as a function of $\chi$. 
 In the case of an incident optical laser with $\omega\sim 1$ eV,
and for a magnetic field of ${\mathcal O}(1 \,{\rm Tesla})$, Eq. (\ref{K1}) implies 
that $P(\gamma\to \gamma\gamma)\simeq 4\times 10^{-14}$ for mass and coupling in the PVLAS range \cite{PVLAS}, i.e. 
$m\sim 10^ {-3}$ eV and $\Lambda=10^ {6}$ GeV. For these figures, the characteristic magnetic length $L$ will be 
macroscopic, i.e. 
$L\simeq 0.5$ m. For a Nd:Yag laser with average power of $1$ Watt 
(and typical wavelength of $\lambda \simeq \mu{\mathrm m}$), we will have that
 $N_{\gamma}(t) \simeq 10^{18}/s$. Consequently, the number of photon splittings per second will be  
$N_{\gamma\gamma} \simeq 4\times10^4$ which is potentially observable.
\begin{figure}[tpb]
\dofigureb{3.1in}{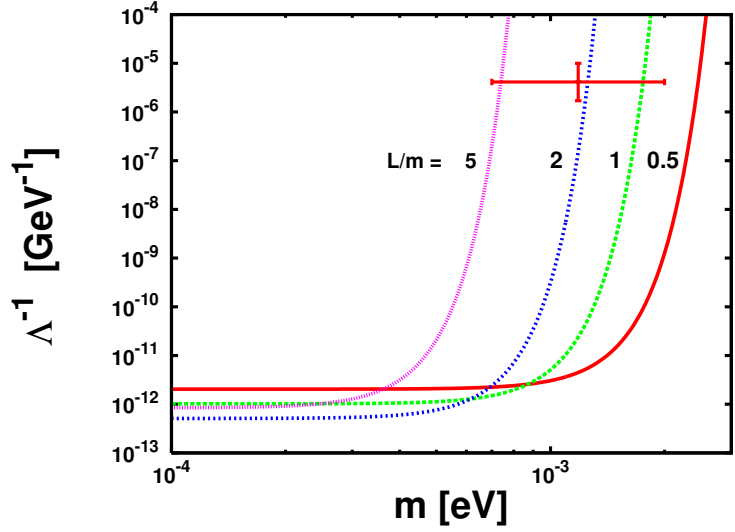}
\caption{\small Exclusion regions for photon splitting 
(above the lines)  at 95\% C.L.
in the $(m~,~\Lambda^{-1})$ plane, for an optical beam of frequency 
$\omega=1$ eV, and a magnetic field ${\rm B}=1$ T of different length (L), 
namely L/m=0.5,~1,~2, and 5. The point 
with error bars corresponds to the PVLAS range. The integrated time 
is $10^{7}$ sec.}
\label{Figure2}
\end{figure}
As it can be appreciated from Fig. \ref{Figure2} the PVLAS region can be 
excluded, at 95 \% C.L. ($10^{7}$ sec of integration time), 
for magnetic inhomogeneity scales smaller than about a meter.

Different assumptions on the magnetic field profiles lead, 
up to numerical factors ${\cal O}(1)$, to the same results. 
This has been cross-checked with a variety of profiles where the magnetic
field vanishes at infinity. Moreover, the different shapes of the magnetic field will
modify the form of ${\mathcal F}(\chi)$ and will therefore change the suppression 
of the  probability when the magnetic inhomogeneity scale is larger than the inverse 
of the resonant momentum $\overline{p}$.

It is instructive to discuss briefly the  massless limit 
of Eq.(\ref{K1}). In this limit, as a consequence of the Breit-Wigner distribution (see e.g.  Eq.(\ref{Pgg}) )
the resonant pole is absent and  
the photon splitting probability is discontinuous in the limit $m\to 0 $.
In the massless case, the total probability of photon splitting should then
be calculated by setting $m=0$ inside the integrand of Eq.(\ref{Pgg}) and the final result can be expressed as
\bea
&& P(\gamma\to \gamma\gamma)=\frac{{\rm B}^2}{256\,\pi\,\Lambda^4}
\, {\mathcal I}_{3}(\xi),
\label{Pm0}\\
&& {\mathcal I}_{3}(\xi)=\xi^2
\int_0^{1/2}\, dx\left(\int_0^{4x(1-x)}\,dy+\int_0^{2x}\,dy\right)
\frac{e^{-\xi^2(1-\sqrt{1-y})^2/2}}{1-y},
\eea
where $\xi=\omega\, L$. If, as previously assumed,  
$\omega=1$ eV and $L \simeq {\mathcal O}(\mathrm{m})$, we
have $\xi\gg 1$, and, in this limit, 
${\mathcal I}_{3}(\xi)\simeq \xi\, \sqrt{2\pi}$. This occurrence implies then
\bea
P(\gamma\to \gamma\gamma)\simeq 
\frac{{\rm B}^2\,\omega\,L}{128\,\sqrt{2\pi}\,\Lambda^4}\, ,
\eea
which is very suppressed due to the fact that the
resonant effect has disappeared for $m =0$.  
As a useful comparison we can also 
report the probability of conversion of a photon into a (massless) 
pseudo-scalar which turns out to be 
$P(\gamma\to \varphi_{P}) = {\rm B}^2 L^2\, \pi/(8 \Lambda^2)$. This 
result follows from the same external magnetized background 
assumed in Eq. (\ref{Bshape}).

Of course, the aforementioned discontinuity is present if dispersion 
effects, associated with the self-energy corrections due to the 
interaction with the external field, are neglected.
These effects typically cause a shift of 
the squared mass, i.e. $m^2\to \overline{m}^2=(m^2+\sqrt{m^2+4m_{\rm B}^2})/2$,
where $m_{\rm B} = \sqrt{{\rm B}\omega/\Lambda}$. 
Corrections, of order ${\cal O}(m_{\rm B})$, arise as well for the 
photon self-energy.
If the mass is in the PVLAS range ($m_{\rm PVLAS}
 \sim 10^{-3}{\rm eV}$) for 
Tesla magnetic fields and couplings $\Lambda \sim 10^6$ GeV, 
$m_{\rm B} \simeq 10^{-6} \, {\rm eV} \ll 
m_{\rm PVLAS}$ and dispersion effects are negligible for resonant photon
splitting.

As already mentioned PVLAS results lack independent confirmations.
In the present paper it has been shown that, 
for the PVLAS range of masses and couplings,  an observable rate of photon 
splittings is expected 
when the external magnetic field can absorb three-momentum in a continuous way.
Therefore, to confirm, in an independent channel,  the PVLAS findings, 
resonant photon splitting represents an intriguing option.

\end{document}